# A Metadata Registry from Vocabularies Up: The NSDL Registry Project

Diane I. Hillmann
Cornell University
Tel: +1 607 387-9207
dih1@cornell.edu

Stuart A. Sutton
University of Washington
Tel: +1 206 228-6709
sasutton@u.washington.edu

Jon Phipps
Cornell University
Tel: +1 607 785-3224
jp298@cornell.edu

Ryan Laundry
University of Washington
Tel: +52 312 316 1000
rjlaundr@u.washington.edu

**Abstract:**
*The NSDL Metadata Registry is designed to provide humans and machines with the means to discover, create, access and manage metadata schemes, schemas, application profiles, crosswalks and concept mappings. This paper describes the general goals and architecture of the NSDL Metadata Registry as well as issues encountered during the first year of the project's implementation.*

## 1. Introduction

In this paper, we describe progress on the development of the National Science Digital Library (NSDL) Metadata Registry (hereafter Registry) as a fundamental piece of core technical architecture. It is not the purpose of this paper to chronicle the short history of research in the area of Web-based metadata registries. For current explications of an array of registry initiatives, see Wagner and Weibel (2005) and Kotok (2003). Needless to say, registries have been a part of the metadata discussions for a number of years, as the need for enabling infrastructure for the Semantic Web has become more critical.

The NSDL Registry will make possible: (1) the unambiguous *identification* of metadata schemas (attribute spaces or element/property sets) and schemes (value spaces or controlled vocabularies); (2) the machine *declaration* for encoding and network transmission of those schemes and schemas; and (3) the *publication* of those schemes and schemas to communities and applications. As part of its core services, the Registry will provide machine-addressable crosswalks and other mappings that relate member terms in the schemes and schemas it contains one to another. In addition, the project will provide well-documented means for individual NSDL projects and others to *identify*, *declare* and *publish* their local schemes and schemas through the Registry. Thus, the Registry will support the key goals of metadata *discovery*, *reuse*, *standardization* and *interoperability*.

The NSDL Registry work is grounded solidly in the NSDL projects facing challenges in the effective deployment of their metadata schemes and schemas. In the past few years, a community of interest within NSDL has emerged. Communication and work among this community has been supported by the proposers through NSDL Communication Portal discussion lists and an NSF/NSDL-sponsored Vocabulary Workshop. Use cases to guide Registry development have been vetted through this community of interest. The community will also assist the project through iterative evaluation during the project's second year.

One of the goals of the NSDL Registry is to provide a stable home for schemes, schemas and application profiles used in the NSDL that lack a maintenance organization with the interest and resources for their long-term maintenance. Another goal is to interact with registries external to NSDL that manage schemes and schemas of interest to the community. It is fundamental to the stability of knowledge organization systems and schemas that their maintenance and evolution be managed as near their source—their promulgating agency—as possible. Thus, while it is meaningful to develop a centralized NSDL Registry, it can only function effectively if it can interact with registries operated by the promulgating agencies just noted. Therefore, we will build on the Web Services currently deployed to address this critical need to provide inter-registry interactions for both humans and machines.

As a result of this need for the NSDL Registry to interoperate with other metadata registries, we define two classes of entities requiring different levels of "management." The first class is made up of those entities *hosted* by the NSDL Registry. These are entities for which the canonical versions reside within the Registry. The Registry provides the promulgators of this class of entity with capabilities to upload and import into the Registry fully-formed entities or to create, edit, and version entities and their content using Registry tools.

The second class of concern to the Registry is *non-hosted* entities. The goal with non-hosted entities is to interact with the registries in which they reside and to expose those entities through the Registry interface. The Registry has no means to "manage" such entities or their content and will limit the functionality offered to discovery and exposure.

## 2. NSDL Registry Services

In essence, the Registry will manage the following hosted top-level entities and their content:
- *Schemas*. Entities that define elements or properties in attribute space namespaces;
- *Schemes*. Entities that define concepts in value space namespaces;
- *Application Profiles*. Entities that provide the means for selecting terms from disparate attribute and value spaces and defining their usage for a specific discourse or practice community (see, Heery & Patel, 2000);
- *Crosswalks*. Entities that define relationships among elements or properties in disparate attribute spaces; and
- *Mappings*. Entities that define relationships among concepts in disparate attribute spaces.

The relationships among these top-level entities are illustrated in Figure 1.

To date, most research implementations for the Web have approached registry research and implementation as a means for managing and promoting reuse of attribute spaces — i.e., the left-hand side of Figure 1. While the NSDL Registry will also be handling attribute spaces, the initial work has focused instead on value space issues—taking on some of these issues at the most granular level and attempting to address the big question: "What should these registries do with knowledge organization systems (KOS) such as thesauri, taxonomies, simple term lists and ontologies and how should such registries operate in an open services environment?" Because controlled vocabularies tend to be more volatile and change is a necessary part of the management challenge, we believe that starting with value spaces will ensure that the decisions we make and the processes we design will work well for less volatile resources.

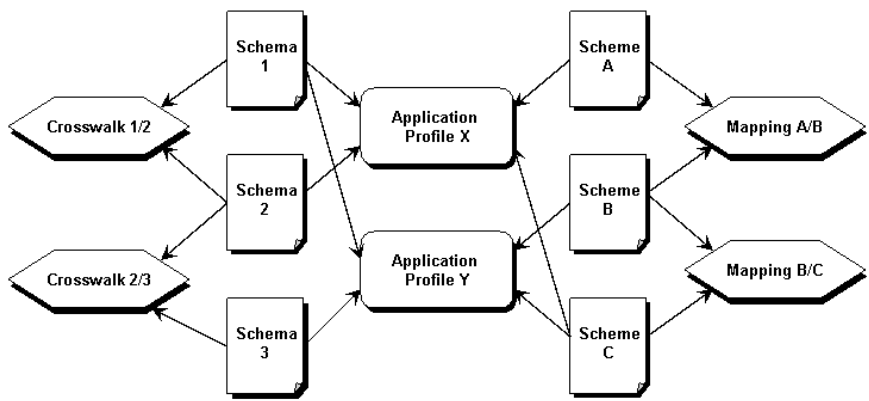

**Figure 1: Top-Level Entities**

It is clear that one measure of the long-term success of the NSDL Registry will be the level of technical transparency of its underlying metadata abstract models and their associated encodings in schema languages. It asks too much of a collection holder or community wishing to develop an application profile to master a schema language in order to generate an appropriate schema. Placing tools in the hands of users that provide the means to generate schemas for submission through simple interface mechanisms, drawing on elements already in existence in the Registry, encourages the use of application profiles and makes them easier for others to discover. In addition, providing a simple means for extending existing schemas to include local elements is also required and will be possible through the schema generation tool.

*2.1. Registry Services for Vocabulary Users*

Although registries have long been regarded as one of the missing parts of web infrastructure, it does not follow that "build it and they will come" is sufficient to persuade either vocabulary owners or users to interact with a registry. Incentives in the form of easily understood value-added services are the key to bringing both owners and users into the Registry—and keeping them coming back.

Registry services can be categorized at the most basic level by whether the initial user of the service is a human or a machine. For human users, initial services in the Registry will be resource discovery and maintenance. A typical use case for human users begins with the need to search or browse the Registry for vocabularies that might suit the needs of a project or community, most often during planning phases for projects or application profiles. Users at this stage are presumed to be looking for a rich and comprehensive result set, which can allow them to explore the range and depth of vocabularies available through the Registry.

For users who have already made a choice, or for whom a choice is determined by community requirements, the Registry will provide services that will allow for the optimal maintenance of chosen vocabularies within instance data. Because one criteria Bruce and Hillmann (2004) assert as a measure of quality of metadata is the currency of the controlled vocabulary terms, a range of services will be offered to assist in keeping vocabularies in applications and instance data current.

Users of particular vocabularies will be able to register their usage and sign up for regular, configurable notification of changes in the vocabularies they use. Notifications can include a variety of options ranging from files that can be used directly in update routines, to human readable change listings that staff can use to update data using established manual processes. Because the goal is to support the maintenance of metadata, Registry developers will work closely with early users to ensure that the array of services offered meet the needs of projects and data providers.

We recognize that the initial categorization of human and machine users breaks down rather quickly, as some of the service components selected by humans are intended for automated provision, but we need to be flexible about how the services are delivered, given the necessity to meet the needs of users at all stages of automated capability.

*2.2. Registry Services for Vocabulary Owners*

Ultimately, registry success relies much more on services to vocabulary owners than it does to other users. If vocabulary owners can't find a reason to continue to update their vocabularies in the Registry, users will need to find other ways outside the Registry to maintain their data or not maintain it at all. Given that reality, it is obviously critical to this category of services to make the Registry an integral part of the document/publish strategy for vocabulary owners and managers, and not just another task with little or no immediate payback.

The first interaction vocabulary owners will have with the Registry is as a user, registering an organization or individual as an agent and registering additional contacts for the agent. From there they provide basic information about the vocabularies they own and/or manage, either as an individual or on behalf of an organization, and designate contacts as maintainers of each vocabulary. This process provides the basis for a continuing relationship between the Registry and the vocabulary, and focuses on setting up properly scoped contact information that can be used for ongoing notification and interaction.

*2.3. Uploading Existing KOS to the Registry*

We consider it likely that in many instances, vocabulary owners will initially continue to manage and update their vocabularies using whatever processes and applications that have served them in the past. Eventually, our goal is to be able to supply services within the Registry that will allow vocabulary owners to shift their maintenance activities to within the Registry, relying on easy, configurable output mechanisms to update vocabulary usage within their own applications and data processes.

In order to support migration of existing vocabularies to the hosted registry management infrastructure, the Registry will provide a flexible KOS upload and import process. This process will support the import of existing KOS from a number of different file formats, including non-XML/RDF formats where the requirements of the vocabulary allow for it. Once the vocabulary has been imported, vocabulary owners and maintainers may request export of the vocabulary in any of the input or output formats that the Registry supports, bearing in mind the potential for data loss with non-XML/RDF formats. Web services will also be provided that will support remote vocabulary maintenance and interaction.

*2.4. Generating KOS within the Registry*

As we noted earlier, one of the goals of the project is to provide developers and maintainers of KOS with the means to author and update those KOS within the Registry environment. While we are committed to being as open as possible in terms of encodings for existing KOS imported into the Registry, by necessity we must be more selective in the scheme authoring environment we implement. Initially we will be developing an editor and validator conforming to the Simple Knowledge Organization System (SKOS) (http://www.w3.org/TR/2005/WD-swbp-skos-core-spec-20051102/). Where possible, we will build on existing work in this area—see, for example the W3C work on SKOS validation (http://www.w3.org/2004/02/skos/core/validation).

Framing the Registry's built-in authoring environment on the evolving SKOS is not without its problems. Currently, there is no direct support in SKOS for handling versioning of KOS concepts. From the beginning of the project, we recognized the absolute need to manage versioning of schemes and schemas as well as their member concepts and terms. It is to these issues that we now turn. We will return to the current limitations of SKOS near the end of the paper.

**3. Versioning Challenges**

Tracking changes in resources is an essential task of a registry. Users need to be able to manage change either by relying on a particular version of a schema or scheme until a particular change makes reconciliation a necessity, or alternatively, by automatically updating to match each new change. The

Registry must support them in carrying out either strategy.

Controlled vocabulary versioning issues occur with both URIs and descriptions. Each can change at two levels: at the term level, where each term change may invoke a change management policy, and at the overall vocabulary level, which is intrinsically different each time a term changes. Because it's not entirely clear what end users of vocabularies will require from registered vocabularies, the Registry will make available historical changes and versions of the vocabularies and individual terms to the extent possible.

The Registry strategy for tracking change relies partially on the software model, where recognition of "diffs" or differences between one version and the next (including who made the changes) are the norm. Use of this model allows a complete history of all changes (and who made the change) to be maintained and accessed by administrators, maintainers and users.

But not all change is important in the evolution and proper usage of vocabularies and terms, and flooding users with undigested information is clearly not an acceptable solution. Based on an in-depth analysis of possible semantic changes and their implications, the Registry will track semantically significant changes to individual terms in ways that will assist users in maintaining their vocabularies and their metadata appropriately.

Because there are distinct differences in the control the Registry has over hosted and non-hosted vocabularies, the Registry policies for each will be separately addressed.

*3.1. General Assumptions:*

1. URIs will remain stable as long as the semantics of the concept do not change;
2. URIs of individual concepts won't contain version information;
3. The Registry must be allow people/services to create dependencies on an identifiable snapshot of a particular representation of a vocabulary and it's relationships;
4. An identifiable snapshot must include the version designation (either "number" or "date");
5. Once published, individual concepts in a vocabulary may be created, updated, or deprecated, but not deleted;
6. Namespaces of vocabulary schemas won't be versioned; and
7. Schema name versioning will only change if the version change would harm backward compatibility

*3.2. URI Changes*

Stability and reliability of concept URIs is critical to the Registry. Determining unambiguously when a maintainer of a hosted term intends to change its semantics will be a challenge with some forms of controlled vocabularies. If the Registry allows registration of simple term lists, without hierarchies or definitions to determine term boundaries, there is no ability to automatically signal any semantic change beyond the addition and deprecation of terms. Mappings between simple term lists and other schemes, or assertions of relationships between undefined terms are also problematic in this context.

Most changes in description of the term, including most changes of definitions and simple additions or changes in term relationships, should not qualify as semantic changes requiring a change in a term URI. In general, non-semantically significant changes might include:

1. Additions of broader, narrower or related terms, when no change in hierarchical placement is made;
2. Changes in definition for clarification, correction of typos or grammar, etc.;
3. Addition of definition or scope note when none is present;
4. Change in term status; and
5. Addition of other information (references, etc.).

Semantic changes, requiring a change in URI, might include:

1. Some instances of term splitting or consolidation;
2. Changes in definition that change the semantics of the term; and
3. Changes in hierarchical relationships, when there is no definition and the hierarchy placement is the only semantic clue.

Enforcement of this policy is challenging, since the initial decision about whether a change requires a new URI is made by the maintainer (the exception is splits or consolidation, where machine validation is possible). It is possible that a combination of explicit questions to the maintainer before a submission and some monitoring by a Registry administrator (particularly focusing on new maintainers) might decrease chances of semantically significant changes being made without triggering a new URI. This is certainly an area where experience will be instructive.

*3.3. Non-Hosted Vocabularies*

Most of the "control" over externally managed vocabularies, particularly in terms of versioning, will be at a policy level, since the maintenance agency processes will be independent of the Registry. If the Registry is to make available any notion of "versioned copies" for these vocabularies, the versioning information at both the vocabulary and term levels must be exposed to the Registry. Ideally, the Registry will at some point be able to ingest vocabulary "snapshots" (if the maintaining agency makes them available) or create from ingestion of term changes viable "versioned snapshots" for use by other services or organizations.

Registry services may be developed to manage agreements with agencies and ingest processes when terms change externally. The Registry should maintain sequenced copies of the concept schemes to be able to track changes over time and to show these copies to vocabulary users, and potentially use them to provide change notification services similar to those provided for hosted vocabularies.

**4. The Challenge of URIs**

There are at least three possible scenarios envisioned for the assignment of term URIs within the Registry:

1. A vocabulary maintainer submits already assigned URIs with the terms;
2. A vocabulary maintainer submits a domain and URI 'template' with the top-level vocabulary description, so that the Registry can use that information to assign URIs; and
3. A vocabulary maintainer asks the Registry to assign URIs.

In the first case, the owner-submitted URIs can be validated to ensure uniqueness, and to some extent the Registry can monitor for instances where semantic changes might require a new URI, but should be able to assume that the vocabulary maintainer is taking responsibility for URI assignments for new terms. In the second instance, the maintainer may not already have assigned URIs, but since they are required in the Registry, a domain can be submitted, along with a decision on whether the term name or a numeric value will be used to create a unique URI, and the Registry can complete the process of assignment when the terms are added. In the last instance, the vocabulary maintainer asks the Registry to assign a URI and the Registry assigns a permanent URI constructed from a base domain (either a domain supplied by the vocabulary owner, or the Registry's native domain), a unique token assigned by the vocabulary owner to the vocabulary itself, and a numeric value assigned to each vocabulary concept. This construct will ensure the uniqueness of each URI and provide support for the W3C Semantic Working Group's "Best Practices Recipes for Publishing RDF Vocabularies" http://www.w3.org/2001/sw/BestPractices/VM/http-examples/).

As part of the effort to analyze the implications of vocabulary changes on the Registry,

it became clear that using term names or labels as part of a URI (a practice common in schema registries, including the DCMI registry) in an effort to improve the "human readability" of URIs, could eventually degrade, particularly given the greater volatility of controlled vocabularies over attribute sets. This would tend to happen particularly in cases where a prefLabel and an altLabel for a concept might be interchanged, for instance when term usage changed over time. For this reason, the Registry will use numeric concept identifiers, as noted above, as a default, and encourage vocabularies that have not already committed to using term names as identifiers to follow suit.

## 5. Notifications, Outputs, and Other Interactions

Like most digital library services, the Registry is designed to operate with the least possible human intervention. For that reason, considerable effort will be devoted to designing and implementing automated notifications that can be easily understood by users, and to which there is adequate support for an appropriate response. Where possible, requests that require simple "yes/no" responses will include clickable links, similar to those now common for email confirmations when registering for discussion lists and other services. In other cases, links to logs, documentation, or specific terms or interactions will be included to assist the users in solving problems that have been the cause of the notification. Vocabulary maintainers will also be prompted to review and resolve identified problems when they log in to the Registry.

Registered users will be able to subscribe to a notification service that will let them know, via Atom/RSS/RDF feed or email, of changes to all or selected vocabularies. Additionally, vocabulary owners may request that routine notifications be sent when:

- registered maintainers have modified terms or term relationships;
- file uploads or service interactions have validation errors or require confirmation (for instance, to confirm whether a term change might qualify as a semantic change requiring a new term); and
- new terms have been added and a new term URI has been created.

Because most Registry interactions with vocabulary owners and maintainers will be in the form of automated notifications, we recognize that creating notifications that are understandable and easily actionable by a broad range of agents will be an enormous challenge. A helpdesk system to track and manage interactions arising from notifications will be essential to the project, as will a full range of supporting documentation.

As part of the enticement for vocabulary owner participation, we anticipate notifying owners when users register their intention to use their vocabularies, providing an incentive to continue maintaining via the Registry system and perhaps also encouragement to continue investing in vocabulary development. This registration of usage is integral to both vocabulary owners and users—each has a strong interest in the participation and activities of the other, and building on that interest will be more likely to contribute to the growth of the Registry than broad appeals to the "common good." Detailed specifications for output formats and mechanisms are still incomplete, but will be an important priority as implementation progresses.

Another reason for broad notification is to prevent nefarious activity within the Registry, without the introduction of extensive security measures that complicate interaction. In instances where a person is maintaining a vocabulary on behalf of another person or organization, notifications to other contacts with interests in the vocabulary provides extra security for the Registry.

### 5.1. Inter-Registry Services

If the vision of distributed registries is to become reality, services between registries must be part of the planning package. Given the expected volatility of some vocabularies, these services must be based on standardized service models and require as little human intervention as possible.

A distributed registry system should allow users to discover schemas, vocabularies and application profiles across the system, without having to "shop" individual registries for an appropriate result. Given the problems of federation-based "metasearch" solutions in the library world, it is unlikely that discovery services in the Registry world could acceptably operate with discovery required to navigate federated "silos." Thus, the Registry will provide APIs that support the interchange of data between metadata registries. Any metadata registry or other service that supports the same APIs will be able to exchange data with the Registry.

**6. SKOS Sufficiency—"Mind the Gap"**

Like Dublin Core, SKOS contains little in the way of guidance or support for meta-metadata, leaving most decisions to the implementer. This is particularly an issue when management of change and versioning is considered. As Tennis (2005) points out in a recent paper, there are basically two methods for concept scheme revision in SKOS: notes and OWL versioning. He suggests some additional extensions to address concept "lumping" or combination of terms as well as concept refinement.

Another issue that SKOS addresses only in its internal documentation is "status." SKOS terms themselves each have a "status"—defined by a small vocabulary of status terms—but the status of terms within a vocabulary cannot be described using SKOS. To some extent this gap in attention to administrative metadata mirrors Dublin Core, which relies exclusively on external standards (like OAI-PMH) to supply the administrative "wrapper" around resource metadata. The Registry will define and support a vocabulary of status terms (registered of course) intended to provide vocabulary users with an indication of whether a term has simply been proposed, is approved (or not), or has been depredated.

While additional support for revision and change management is welcome, extensions that address only the "human-friendly" aspects of concept management provide only a partial solution. The Registry software will, as a default, track every change made to concepts, and presenting this history of change to users without extensive editing by humans will be necessary, if not necessarily simple. Reliance on human-created and maintained notes to present change history to users is not a scalable solution for a registry that must rely as much as possible on automated processes. Many of the maintainers of vocabularies interacting with the Registry will not be trained in vocabulary management, so expectations that they will understand SKOS or thesaurus concepts sufficiently to construct standard notes are probably misplaced.

It is also possible that some flavors of output desired by users will require distribution of the full change histories maintained by the Registry, which suggests a need for standardized methods for capture, characterization and exposure of machine-created and readable concept changes. Other management information, like "status" might also be included in some desired output.

**7. Conclusion**

Building a registry from the most granular pieces "up" to more general, aggregated expressions provides both important opportunities and significant potential for stumbles. Without the development of SKOS, it would clearly not be feasible, and given that there have not, at this writing, been significant SKOS implementations, there are still a few leaps of faith required. One interesting question it's still too early to answer is: how will experience building this end of the Registry inform the other parts? Each phase implies a shift in focus, and a consolidation of lessons, but each builds significantly on the next.

This material is based upon work supported by the National Science Foundation under Grant No. DUE-0532828. Any opinions, findings, and conclusions or recommendations expressed in this material are those of the author(s) and do not necessarily reflect the views of the National Science Foundation.